\documentclass[11pt,twoside]{article}
\usepackage[utf8]{inputenc}
\usepackage{amssymb}
\usepackage{amsfonts}
\usepackage{amsmath}
\usepackage[english]{babel}
\usepackage[pdftex]{graphicx}
\usepackage{fancyhdr}
\usepackage[normalem]{ulem}

\begin{document}

\fancypagestyle{plain}{%
\fancyhf{}%
\fancyhead[LO, RE]{XXXVIII International Symposium on Physics in Collision, \\ Bogot\'a, Colombia, 11-15 September 2018}}

\fancyhead{}%
\fancyhead[LO, RE]{XXXVIII International Symposium on Physics in Collision, \\ Bogot\'a, Colombia, 11-15 September 2018}

\title{Searching for Heavy Neutral Leptons\\with Displaced Vertices at the LHC}
\author{%
Asmaa Abada$\thanks{e-mail: abada@th.u-psud.fr}$, 
Xabier Marcano$\thanks{e-mail: xabier.marcano@th.u-psud.fr}$,
\\
Laboratoire de Physique Théorique, CNRS\\
Univ. Paris-Sud, Université Paris-Saclay, 91405 Orsay, France
\\ \\
\uline{Nicolás Bernal}$\thanks{e-mail: nicolas.bernal@uan.edu.co - speaker}$ and 
Marta Losada$\thanks{e-mail: malosada@uan.edu.co}$
\\
Centro de Investigaciones, Universidad Antonio Nariño\\
Carrera 3 Este \# 47A-15, Bogotá, Colombia
}
\date{}
\maketitle

\begin{abstract}
Heavy Neutral Leptons are naturally present in many well-motivated extensions of the Standard Model.
If their mass is of few dozens of GeVs, they can be long-lived and lead to events with displaced vertices, giving rise to promising signatures due to the low background.
We revisit the opportunities offered by the LHC to discover these long-lived states via searches with displaced vertices.
We study in particular the implication on the parameter space sensitivity when all mixings to active flavors are taken into account. 
\end{abstract}

\section{Introduction}
Longstanding challenges of the Standard Model (SM) like the origin of neutrino masses, the nature of dark matter and the generation of the observed baryon asymmetry of the Universe may be related to the existence of heavy neutral leptons (HNL), sterile from the SM gauge point of view~\cite{Abazajian:2012ys}.
Their mass scale is a priori not fixed, giving rise to a very rich phenomenology.
For instance, if they are very light, they can affect neutrino oscillations, as suggested by some neutrino oscillation anomalies.
In contrast, if they are very heavy, experiments at the high intensity frontier can probe their low-energy imprints, for instance in lepton flavor violating processes.

In this work we focus on collider searches for the HNL, in particular in the mass range up to a few tens of GeV.
For the sake of generality, we will not invoke a particular link to any neutrino mass generation mechanism.
Interestingly, in this mass ballpark the HNL could be long-lived and travel a macroscopical distance before decaying inside the (LHC) detectors.
In such situation, dedicated searches looking for displaced vertices (DV) are needed to exploit such a distinctive signature, with a particularly low SM background.

In this article we first present the framework used, based on a simplified $3+1$ model where the SM particle content is enhanced with a single HNL.
Then we discuss the conditions required to have a detectable DV, emphasizing on the role played by the flavor dependence.
Finally, we present in detail the LHC sensitivities for the dimuon channel, paying particular attention to event-triggering and potential backgrounds.
Further details can be found in Refs.~\cite{Abada:2018sfh,Marcano:2018fto}.

\section{A Simplified 3+1 Model}
In this work we minimally extend the SM by adding a single sterile neutrino, the HNL.
Because we are mainly interested in the collider phenomenology of the HNL, we do not assume any specific underlying mechanism for the active neutrino mass generation.
This means that we are not demanding the new sterile state to play a role in the generation of active neutrino masses, and hence the HNL mass $m_N$, and its mixings with the light neutrinos, $V_{\ell N}$ with $\ell=e$, $\mu$, $\tau$, are free independent parameters.
These mixings define the interaction strength of the HNL via charged currents, as well as the neutral currents to the $Z$ and $H$ bosons.
The unitary $4\times4$ lepton mixing matrix is
\begin{equation}
U_\nu^{3+1} = \left(\begin{array}{cc}
\tilde U_{\rm PMNS} & V_{\ell N}\\
V_{N\ell} & U_{NN}
\end{array}
\right)\,,
\end{equation}
where the $3\times 3$ $\tilde U_{\rm PMNS}$ matrix is similar to the usual PMNS matrix up to small non-unitarity corrections due to the presence of light-sterile $V_{\ell N}$ mixings.

In our numerical simulations, we use {\tt MadGraph5\_aMC@NLO}~\cite{Alwall:2014hca}  to generate  the HNL production events from proton-proton collisions and  {\tt Pythia 8.2}~\cite{Sjostrand:2014zea} for its subsequent decay and treatment of the DV.
Finally, Les Houches Event files are obtained using {\tt MadAnalysis5}~\cite{Conte:2012fm}.
Jets are reconstructed using an anti-$k_T$ algorithm with a radius of 0.4 and minimum $p_T$ of 5~GeV.
The HNL total decay width and branching ratios were however computed analytically and used as inputs for {\tt MadGraph5\_aMC@NLO},  using the {\tt time\_of\_flight} option to generate the events with DV.

\section{Displaced Vertices and Flavor Dependence}
In order to have a detectable DV, the total decay length of the HNL must be of the same order of the size of the detector, which in the case of ATLAS or CMS is between 1~mm and 1~m from the interaction point.
Interestingly, the region with decay lengths relevant for DV searches at the LHC lies in the few GeV region, where present experimental constraints on the mixing angles are weaker.
The total width  can be approximated as
\begin{equation}\label{eq:Nwidth}
\Gamma_N\propto G_F^2\, m_N^5 \sum_{\ell=e,\,\mu,\,\tau} \big| V_{\ell N} \big|^2\,,
\end{equation}
above the tau mass threshold.

In collider searches for HNL, the simplified hypothesis where the HNL mixes only to one flavor at a time is usually assumed.
This is justified in prompt decayed HNL searches, since the number of events with charged leptons of a flavor $\ell$ depends only on the corresponding mixing $V_{\ell N}$.
However, in DV searches, the total decay width plays a crucial role defining where the HNL decays and, since it depends on all the mixings $V_{\ell N}$, cf. Eq.~\eqref{eq:Nwidth}, one cannot conclude independently of each of the mixings.

\begin{figure}[t!]
        \begin{center}
        \includegraphics[width=.49\textwidth]{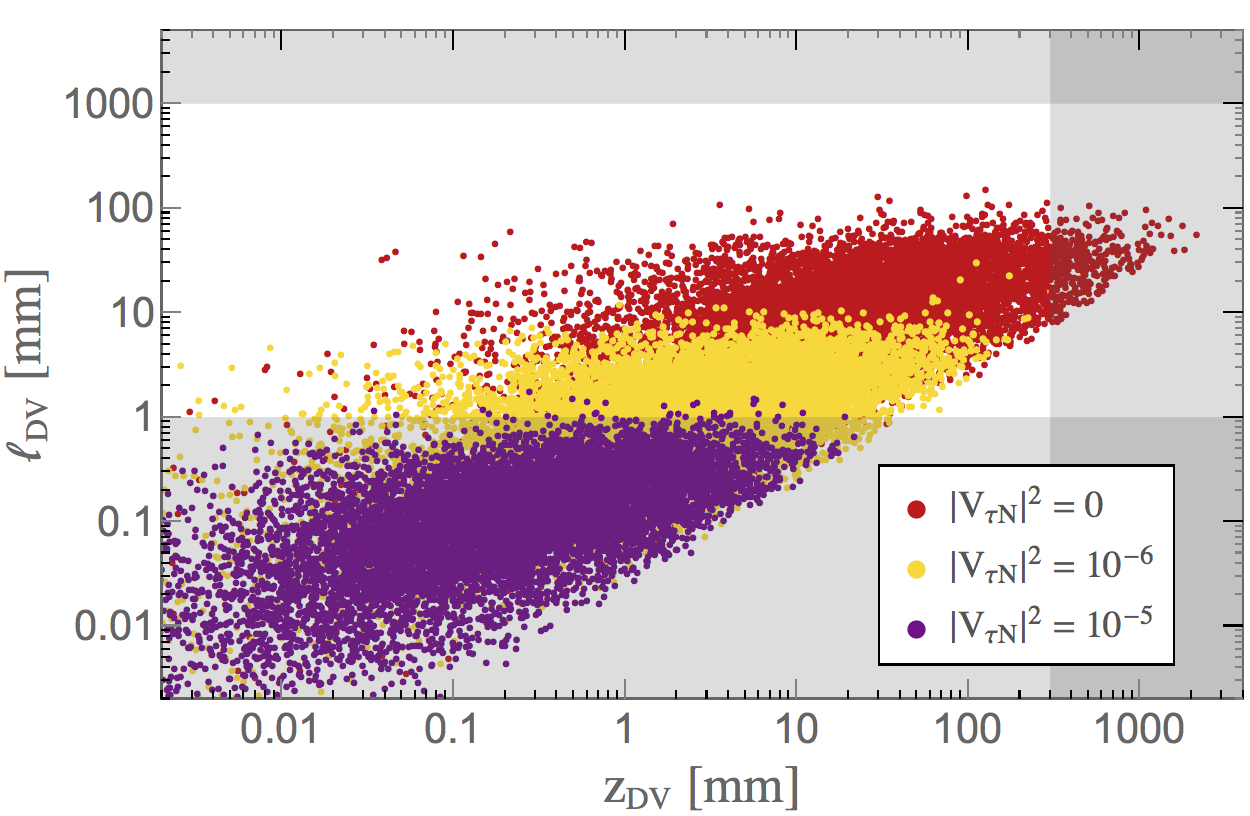}
                \caption{Distributions of the DV coming from a HNL decay in the $(z_\text{DV}, \ell_\text{DV})$ plane. The production channel is $pp\to e^\pm N$, with $p_T^e>25$GeV and $ |\eta^e|<2.5$. We fix $m_N=15$~GeV, $|V_{eN}|^2=10^{-7}$ and $V_{\mu N}=0$ in all the cases, while $|V_{\tau N}|^2= 0$ (red), $10^{-6}$ (yellow) and $10^{-5}$ (purple). As $|V_{\tau N}|^2$ increases, the HNL becomes more prompt and, therefore, insensitive to the DV searches. The white region is the potential  DV area.
        }\label{DVdistributionsVtauN}
        \end{center}
\end{figure}

Fig.~\ref{DVdistributionsVtauN} displays the distributions for the HNL decay position in the $(z_{\rm DV},\,\ell_{\rm DV})$ plane, where $z_{\rm DV}$ is the displacement along the beam axis and $\ell_{\rm DV}$ the displacement in the transverse plane.
We have generated $pp\to e^\pm N$ events with fixed values of $m_N=15$~GeV, $|V_{eN}|^2=10^{-7}$ and $V_{\mu N}=0$, and take the mixing $|V_{\tau N}|^2$ equal to zero (red), $10^{-6}$ (yellow) and $10^{-5}$ (purple).
Since the mass and electron mixing are fixed, the same number of HNL are produced in all  cases.
Nevertheless, the sensitivity of the DV searches to the HNL  depends on the amount of events within the DV fiducial volume, $\ell_{\rm DV} \in [1~{\rm mm},\, 1~{\rm m}]$ and $z_\text{DV}<300$~mm.
In this example, these kinds of searches are very sensitive to the single mixing case, as can be seen in Fig.~\ref{DVdistributionsVtauN}, most of the red points are within this area.
However, increasing values of the mixing $V_{\tau N}$ enhances the total width without affecting the production cross section, shifting the distribution towards lower displacements and thus reducing  the efficiency of the DV searches. Alternatively, some of the cases escaping the detector in the single mixing scenario could lead to DV signatures when all mixings are taken into account.
Consequently, these kinds of searches cannot  explore independently the mixing of each flavor and they should take into account the flavor combination entering in the total width in Eq.~\eqref{eq:Nwidth}.

\section{Dimuon Channel}
\label{sec:sensitivities}
DV signatures for the dimuon channel, where the HNL decays into $N\to\mu^+\mu^-\nu$, have been previously explored in Refs.~\cite{Izaguirre:2015pga,Dube:2017jgo} considering a lepton jet topology and assuming a single mixing scenario.
Here, however, we will closely follow the recent ATLAS analysis in Ref.~\cite{Aaboud:2018jbr}, where this kind of events are triggered by requiring two opposite-sign collimated muons, with transverse momenta larger than 15 and 20~GeV, pseudorapidity $\eta<2.5$ and a small angular separation $\Delta R_{\mu\mu}<0.5$, although lowering the $p_T$ thresholds would increase the signal acceptances~\cite{Abada:2018sfh}.
Additionally, in order to identify the DV we impose 1 mm $<\ell_{{}_{\rm DV}}<1$~m and $z_{{}_{\rm DV}}<300$~mm.
In order to avoid potential background from resonant $Z$ boson production and other events leading to high dimuon invariant masses $m_{\mu\mu}$, we impose a cut on the dimuon invariant mass, $m_{\mu\mu}<m_N$.

The main SM backgrounds for this channel come from low-mass Drell-Yan processes and from single and top pair production.
Other possible sources for the background are cosmic-ray muons, muons with relatively low momentum in multi-jet events, or muons from long-lived mesons, this latter being important at low dimuon invariant masses~\cite{Dube:2017jgo,Aaboud:2018jbr}.
In order to reduce the background from top production and multi-jet processes, we require muons to be isolated from jets~\cite{Aaboud:2018jbr} and that there is a low hadronic activity~\cite{Dube:2017jgo}.
In particular, we ask for an angular separation between any jet and the muons such that
\begin{equation}\label{cutDeltaRmuj}
\Delta R_{\mu j}> {\rm min}\left(0.4,\,0.04 + \frac{10\,{\rm GeV}}{p_T^\mu}\right)\,,
\end{equation}
and also apply a track-based isolation criteria of
\begin{equation}\label{cutID}
I^{\rm ID}_{\Delta R=0.4}\equiv\frac{\sum_{ j} |p_T^{ j}|}{p_T^\mu}<0.05\,,
\end{equation}
where the sum goes over all the jets satisfying $p_T^j>0.5$~GeV and $\Delta R_{\mu j} < 0.4$.
High hadronic activity is vetoed by demanding $H_T<60$~GeV, where $H_T$ is the scalar sum of $p_T$ of all  the jets with $p_T>30$~GeV.

Finally, cosmic-ray background is removed by requiring~\cite{Aaboud:2018jbr}
\begin{equation}\label{cutcosmic}
\sqrt{ \big(\Delta\eta_{\mu\mu}\big)^2 + \big(\pi-\Delta\phi_{\mu\mu}\big)^2} > 0.1\,.
\end{equation}

We will assume that these cuts remove all possible backgrounds of DV coming from SM processes.
Under such hypothesis,
contours corresponding to the $2\sigma$ exclusion reach can be defined by requiring 3 signal events after cuts.

\begin{figure}[t!]
\begin{center}
\includegraphics[width=0.49\textwidth]{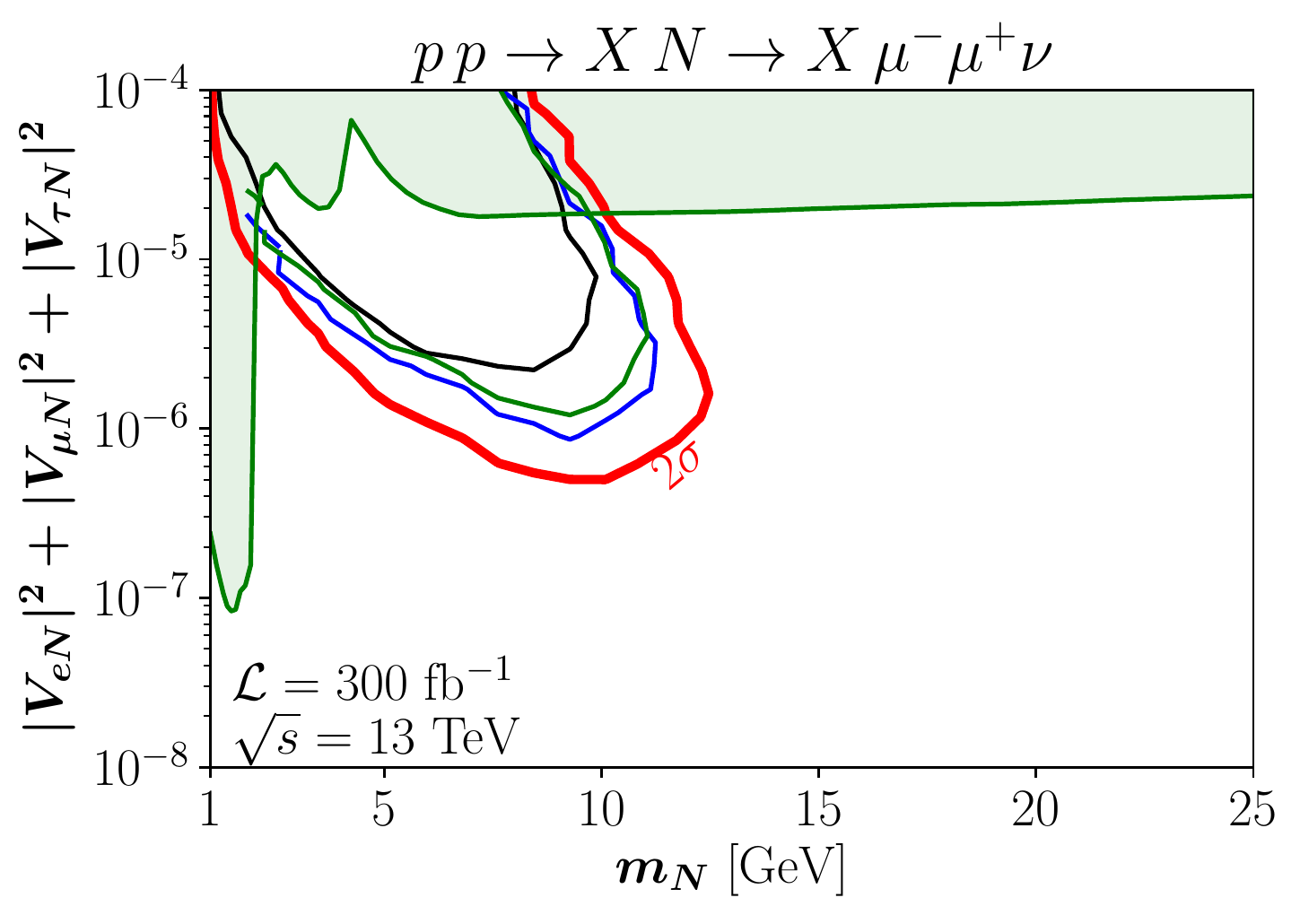}
\includegraphics[width=0.49\textwidth]{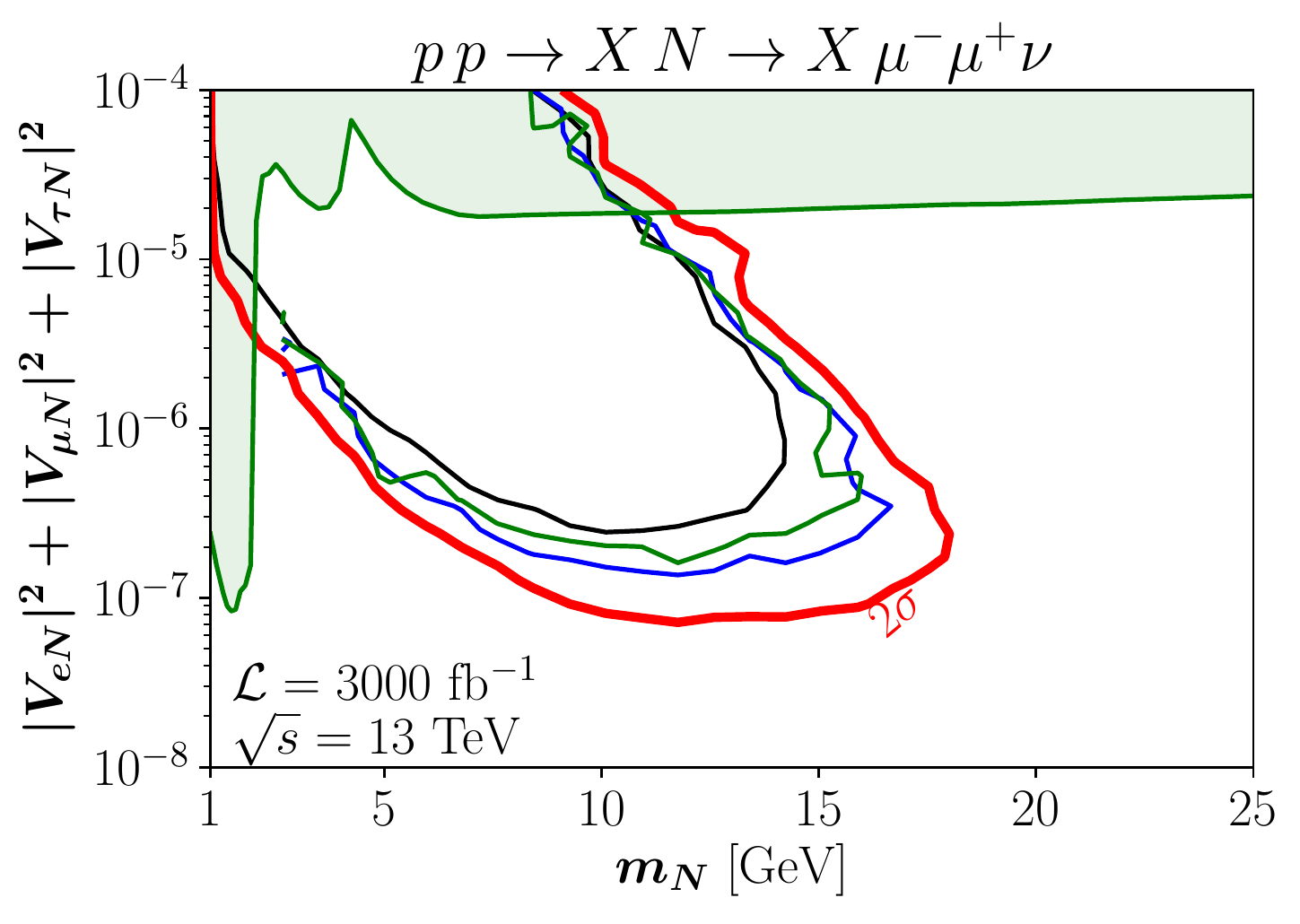}
\caption{Inclusive HNL production followed by its decay $N\to\mu^+\mu^-\nu$, for mixings $|V_{e N}|^2:|V_{\mu N}|^2:|V_{\tau N}|^2$=1:0:0 (black), 1:1:0 (blue), 1:1:1 (green) and 0:1:0 (thick red).
The solid lines are contour lines for $2\sigma$ exclusion with DV at the LHC 13 TeV with $\mathcal L = 300$~fb$^{-1}$ (left) and $\mathcal L = 3000$~fb$^{-1}$ (right) and the cuts described in Section~\ref{sec:sensitivities}.
Shaded green areas are excluded by experimental bounds.
}\label{DVevents-inclusive-ATLAS}
\end{center}
\end{figure}

We show in Fig.~\ref{DVevents-inclusive-ATLAS} the potential LHC sensitivity after collecting $300~{\rm fb}^{-1}$ (left panel) and $3000$~fb$^{-1}$ (right panel).
Different flavor hypothesis are considered in different colors, $|V_{eN}|^2:|V_{\mu N}|^2:|V_{\tau N}|^2=$1:0:0 (black), 1:1:0 (blue), 1:1:1 (green) and 0:1:0 (thick red).
It is clear that the LHC with $\mathcal{L}=300~(3000)$~fb$^{-1}$ luminosity could further probe the parameter space up to one (two) order (orders) of magnitude  below  present bounds.
The exact sensitivity reach in mass and mixings depends however on the chosen HNL flavor pattern and, therefore, we stress again on the importance of exploring all possible final flavor channels.

\section{Conclusions}
Heavy neutral leptons (HNL) naturally appear in many well-motivated scenarios, and could play a key role in the solution of many Standard Model problems.
If their masses are in the few GeV ballpark they can lead to displaced vertices at the LHC, a particularly interesting signature due to the low SM backgrounds.
We discussed different production mechanisms, emphasizing the impact of the flavor structure of the HNL.
Additionally, we studied the dimuon decay channel case, and found that a dedicated experimental analysis could improve the bounds by at least one (two) order of magnitude in the sum of the squared mixing angles if $m_N\simeq 3$ to 17~GeV, for integrated luminosities of 300 (3000)~fb$^{-1}$.
We finally pointed out the importance of the complementarity of different channels to explore the full parameter space.

\section*{Acknowledgments}
We acknowledge partial support from the European Union Horizon 2020
research and innovation programme under the Marie Sk{\l}odowska-Curie: RISE
InvisiblesPlus (grant agreement No 690575)  and
the ITN Elusives (grant agreement No 674896).
NB and ML are also supported by the Universidad Antonio Nariño grants 2017239 and 2018204.
NB was also partially supported by the Spanish MINECO under Grant FPA2017-84543-P.

\end{document}